\documentclass[article,nojss]{jss} 

\usepackage{amssymb,bm,amsthm,color,scalerel}
\usepackage{tikz}
\usetikzlibrary{positioning}
\usetikzlibrary{svg.path}
\usepackage{euler}
\usepackage{hyperref}
\usepackage{url}
\definecolor{orcid_color}{HTML}{A6CE39}

\hypersetup{pdfborder={0 0 0}} 
\DeclareRobustCommand{\orcidicon}{%
	\raisebox{.2mm}{\scalerel*{%
	\begin{tikzpicture}[xscale=1,yscale=-1,transform shape]
	\filldraw[color=orcid_color] svg {M256,128c0,70.7-57.3,128-128,128C57.3,256,0,198.7,0,128C0,57.3,57.3,0,128,0C198.7,0,256,57.3,256,128z};
	\filldraw[color=white] svg {M86.3,186.2H70.9V79.1h15.4v48.4V186.2z} svg {M108.9,79.1h41.6c39.6,0,57,28.3,57,53.6c0,27.5-21.5,53.6-56.8,53.6h-41.8V79.1z M124.3,172.4h24.5
		c34.9,0,42.9-26.5,42.9-39.7c0-21.5-13.7-39.7-43.7-39.7h-23.7V172.4z} svg {M88.7,56.8c0,5.5-4.5,10.1-10.1,10.1c-5.6,0-10.1-4.6-10.1-10.1c0-5.6,4.5-10.1,10.1-10.1
		C84.2,46.7,88.7,51.3,88.7,56.8z};
	\end{tikzpicture}}{|}}%
}

\newcommand{\orcid}[1]{\href{https://orcid.org/#1}{\orcidicon}}

\theoremstyle{plain}
\newtheorem{theorem}{Theorem}[section]
\theoremstyle{definition}
\newtheorem{example}[theorem]{Example}

\newcommand{\aut}[1]{Aut(#1)}

\author{Colin Cleveland$^{*}$, Chin-Yen Lee$^{*}$, Shen-Fu Tsai, Wei-Hsuan Yu, Hsuan-Wei Lee$^{\dagger}$ }
\Address{
  Colin Cleveland\\
  Department of Informatics, King's College London, London, UK\\
  E-mail: \email{colin.cleveland@kcl.ac.uk}\\
  Chin-Yen Lee\\
  Department of Mathematics, National Central University, Taoyuan, 32056, Taiwan\\
    E-mail: \email{109281001@cc.ncu.edu.tw}\\
Shen-Fu Tsai\\
  Department of Mathematics, National Central University, Taoyuan, 32056, Taiwan\\
    E-mail: \email{parity@math.ncu.edu.tw}\\
    Wei-Hsuan Yu\\
  Department of Mathematics, National Central University, Taoyuan, 32056, Taiwan\\
    E-mail: \email{whyu@math.ncu.edu.tw}\\
  Hsuan-Wei Lee$^{\dagger}$\\
  Institute of Sociology, Academia Sinica, Taipei City, Taiwan\\
    E-mail: \email{hwwaynelee@gate.sinica.edu.tw}
}

\title{Graphlet and Orbit Computation on Heterogeneous Graphs}

\Abstract{

Many applications, ranging from natural to social sciences, rely on graphlet analysis for the intuitive and meaningful characterization of networks employing micro-level structures as building blocks. However, it has not been thoroughly explored in heterogeneous graphs, which comprise various types of nodes and edges. Finding graphlets and orbits for heterogeneous graphs is difficult because of the heterogeneity and abundance of semantic information. We consider heterogeneous graphs, which can be treated as colored graphs. By applying the canonical label technique, we determine the graph isomorphism problem with multiple states on nodes and edges. With minimal parameters, we build all non-isomorphic graphs and associated orbits. We provide a Python package that can be used to generate orbits for colored directed graphs and determine the frequency of orbit occurrence. Finally, we provide four examples to illustrate the use of the Python package.

}
\Keywords{graph isomorphism, orbit isomorphism, graphlet}

\begin{document}


\def\thefootnote{*}\footnotetext{These authors contributed equally to this work}\def\thefootnote{\arabic{footnote}}
\def\thefootnote{\dagger}\footnotetext{Corresponding author}\def\thefootnote{\arabic{footnote}}

\maketitle

\hypersetup{pdfborder={0 0 1}} 

\section{Introduction}

We live in a data-rich world and it's easy to obtain immense, directed, dynamic, and networked data. Graphlet analysis is a useful analytical method for studying graph-based data by decomposing graphs into many basic and often meaningful elements. Researchers can explain the mechanism of these small unit graphs by focusing on the numbers of these elements and conducting statistical tests and other analyses. This approach has a wide range of applications, including biology \cite{milenkovic2008uncovering}, chemistry \cite{wang2021input}, business \cite{zhou2021competitive}, and social sciences \cite{charbey2019stars, ashford2022understanding}. 

Graphlets are described as small induced subgraphs of a potentially huge network that arise at any frequency. An induced subgraph indicates that after selecting the nodes in the major network, all the edges connecting them must be selected to form the subgraph. Within graphlets, symmetry groups of nodes, known as automorphism orbits (or orbits for brevity), are utilized to characterize the various topological locations that a node might occupy. Orbits have also been utilized to generalize the concept of node degree \cite{prvzulj2007biological, hovcevar2014combinatorial}. If two nodes map to one another in an isomorphic projection of the graphlet onto itself, they belong to the same orbit. To completely decompose a graph of a given size into a certain graphlet and orbit requires good combinatorial mathematics so that nothing is missed. Existing approaches for counting graphlets and orbits rely on direct enumeration; to count them, all network embeddings must be located. Good algorithms are also required to help researchers decompose their networks, number images and nodes, and record the number \cite{hovcevar2014combinatorial, ahmed2017graphlet, ribeiro2021survey}. 

Complex networks are composed of numerous nodes and edges, each containing a large amount of information. Heterogeneity is an inherent trait of heterogeneous graphs, that is, the nodes and edges of diverse types. For instance, different types of nodes have distinct characteristics, and their features may exist in distinct feature spaces. Nodes can have multiple states, such as atoms, gender, and links, as well as multiple states, such as type, direction, and sign. Such heterogeneous networks are pervasive in the natural world, where many types of nodes and edges are detected. This variety of real-world systems is frequently attributed to the fact that the data in applications typically contain semantic information. Researchers have recently used graphlets and orbits with multiple types of nodes or edges in their analyses. For instance, by generalizing the graphlet-based node role concept to directed graphlets, the authors of \cite{sarajlic2016graphlet} conducted sophisticated descriptive and predictive analyses of directed networked data that were not possible with undirected graphlets or other directed network statistics. In addition, by using edge information, \cite{jia2022encoding} demonstrated that edge-type encoded graphlet techniques can outperform the classic graphlet degree vector approach by a large margin in a node classification challenge. \cite{guan2017information} introduced evolutionary characteristics of weighted graphlets to model investors of various categories as a single entity, taking their co-holding strength into account.

To the best of our knowledge, the current graphlet-related algorithms and packages cannot easily handle networks with multiple state nodes and links simultaneously in a systematic manner. Most graphlet packages and algorithms deal with homogeneous graphs. More importantly, the number of possible enumerations grows exponentially if the underlying complex networks are heterogeneous. We hope to develop a suite that can help researchers determine the number of graphlets and orbits they are looking for within a reasonable amount of computation, given the numbers of points, node states, and link states. Additionally, we will help researchers to decompose and number graphlets and orbits within the network. There have been related studies on graphlets on heterogeneous graphs. In \cite{rossi2020heterogeneous, carranza2020higher}, the authors proposed a typed graphlet that generalizes graphlets to heterogeneous graphs, and they outlined a system for counting the occurrences of such typed graphlets and higher-order spectral clustering. However, compared to their work, we focus on enumerating all graphs and orbits and return a vector consisting of the number of occurrences of graphs and orbits of a given graph.

In Section 2, the definitions of colored graphs and orbits are presented. Colors are used to denote various states or types of nodes and edges. We will examine how a colored graph can be used to depict its orbits. Given a heterogeneous graph with $n$ vertices, $v_c$ types of vertices, and $e_c$ types of edges, we calculate the number of graphs and orbits it contains. See Table \ref{tab1} for the numbers of graphs and orbits for a given $(n, v_c, e_c)$. For example, there are 900 graphs and 2,925 orbits of type at most $(n, v_c, e_c)=(4,1,5)$, which means the graphs with 4 vertices, one type of vertex, and at most 5 types of edges. In short, we assign the same graph a fixed label, add new vertices and edges to the small graph, and finally complete the classification. However, because the possible enumeration of graphlets and orbits would soon become huge and our goal is to completely classify the orbits of a colored directed graph, it is not realistic to exhaust all possible enumerations. We focus only on heterogeneous graphs with a reasonably small  parameter set $(n, v_c, e_c)$. However, by converting the problem of counting orbits into solving a linear system of equations, \cite{hovcevar2016computation} showed that one only needs to count the frequency of some special graphs but not all subgraphs. Hence, determining whether two orbits are isomorphic is a graph isomorphic problem, for which there are several effective algorithms based on locating the isomorphism or canonical form. Because the orbit isomorphism problem is an exact graph isomorphism problem, canonical labeling methods are not required. Therefore, the frequency of a subgraph can be determined by computing and comparing the canonical form of each subgraph. One can refer to \cite{melckenbeeck2016algorithm,melckenbeeck2018efficiently,ribeiro2021survey,sarajlic2016graphlet} for methods for solving linear equations. Moreover, \cite{kreher1999combinatorial} provided a robust method for solving canonical forms. We utilize these techniques to obtain the graphlets and orbit information. In Section 3, we present four examples to demonstrate the proposed package. We introduce three main functions that can be used to distinguish graphlet orbits, generate a list of non-isomorphic heterogeneous graphs and orbits, and return a vector consisting of the numbers of occurrences of graphs and orbits of a given graph. Finally, in Section 4, we present our concluding remarks.

\section{Algorithm}
\label{sec: algorithm}
In this section, we first define the \textit{orbits} of heterogeneous graphs. Then we describe our underlying orbit generation algorithm when \code{(Di)Generator} is called to generate all orbits of order up to \code{n}, with number of vertex colors up to \code{sizev} and number of edge colors up to \code{sizee}.\\

\subsection{Orbit definition}
A colored (di-)graph $M$ of order $n$ can be represented by an $n\times n$ matrix $A_M$ with entries from $\mathbb N\cup \{0\}$.
The diagonal and off-diagonal entries represent the colors of the vertices and edges, respectively. Here, colors represent the types of nodes and edges in the heterogeneous graphs. 
Thus, a colored graph is a complete graph $K_n$ with colors for each vertex and edge. We say $M$ is of \emph{type} $(n,v_c,e_c)$ if its order is $n$ and vertices and edges have $v_c$ and $e_c$ colors, respectively. \\

\subsection{Overview}
We first generate all non-isomorphic graphs of order up to \code{n} with up to \code{sizev} vertex colors and \code{sizee} edge colors, as described in Section~\ref{section:graph enumeration}. Then for each of the generated graph $M$ we collect all its associated orbits as described in Section~\ref{section:orbit enumeration from a graph}. Finally we remove duplicate, i.e., isomorphic orbits. We summarize the overall algorithm in the pseudocode below.

\begin{verbatim}
GenerateOrbits(n, sizev, sizee) {
  Let Q be an empty set
  
  Generate Gs, the set of all non-isomorphic graphs
  
  For each G in Gs
    Insert into Q all orbits associated with G

  Remove isomorphic orbits from Q

  return Q
}
\end{verbatim}

\subsection{Graph enumeration}\label{section:graph enumeration}
To enumerate all graphs of order up to \code{n} with up to \code{sizev} vertex colors and \code{sizee} edge colors, we first generate all graphs of order up to 3 with single vertex color and up to 2 edge colors. Then based on that in each of the following \code{(sizee-2)} iterations we enumerate all possible outcomes of adding one edge color. Then based on that in each of the following \code{n-3} iterations we enumerate all possblie outcomes of adding one vertex. Finally we enumerate all possible vertex colorings using up to \code{sizev} colors. We summarize in the pseudocode below. In our implementation, we save as much partial results as possible to files so that successive calls do not cause redundant computation.
\begin{verbatim}
GenerateGraphs(n, sizev, sizee) {
  Let Gs = GenerateGraphs(3, 1, 2)

  For e = 3, 4, ..., sizee
    Let Gt be an empty set
    For each G in Gs
      Insert to Gt all possible result of adding one edge color to G

    Append Gt to Gs
    Deduplicate Gs

  For m = 4, 5, ..., n
    Let Gt be an empty set
    For each G in Gs
      Insert to Gt all possible results of adding one vertex to G

    Append Gt to Gs
    Deduplicate Gs

  Let Gt be an empty set
  For each G in Gs
    Insert to Gt all possible results of coloring vertices of G using up to sizev colors

    Deduplicate Gt

  return Gt
}
\end{verbatim}
\subsection{Orbits enumeration from a graph}\label{section:orbit enumeration from a graph}
For $v\in V$, we denote by $v^g$ the action of $g$ in symmetric group $S_n$ on $V$ (applying the function $g$ to $v$ and obtaining the value $g(v)$).
This action induces a new action on $\mathcal G_n$.
In other words, $M^g$ is the graph obtained by executing action $g$ on $V$.
Two graphs $M_1$ and $M_2$ are said to be isomorphic, denoted by $M_1\simeq M_2$, if there exists a $g\in S_n$ such that
$M_1^g=M_2$.
The automorphism group of graph $M$ is the set of actions $g$ in $S_n$ that produce an identical graph when acting on $M$:
\begin{equation}
    \aut{M} =\left\{ g\in S_n: M^g=M \right\}.
\end{equation}
The graphlet orbit of interest is the orbit of a vertex $v$ of $M\in\mathcal G$ with respect to the automorphism group $\aut{M}$:
\begin{equation}
  \mathcal{O}_{v,M}:=   \{v^g: g\in \aut{M} \}.
\end{equation}
Note that $\mathcal{O}_{w,M}=\mathcal{O}_{v,M}$ for all $w\in\mathcal{O}_{v,M}$. We assign a labeled graph $O_v$ to $\mathcal{O}_{v,M}$ such that the first vertex corresponds to reference point $v$, i.e.,
$O_v=O_w$.

\begin{figure}[ht]
    \centering
    \includegraphics[scale=0.5]{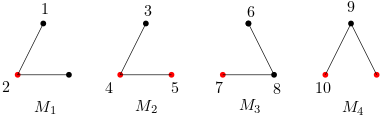}
    \caption{Connected orbits of  type $(n,v_c,e_c) = (3,2,1)$; there are four host graphs and 10 orbits.}
    \label{fig2}
\end{figure}

We list all connected orbits of type $(n,v_c,e_c) = (3,2,1)$ in Figure \ref{fig2}.

\begin{table}
    \centering
    \begin{tabular}{c|c|c|c}
$(v_c,e_c)$\textbackslash $n$& 3        & 4           & 5                     \\
\hline
(1,1)& (4,6)    & (11,20)     & (34,90)                     \\
(1,2)& (10,18)  & (66,165)    & (792,3132)                  \\
(1,3)& (20,40)  & (276,816)   & (10688,48400)               \\
(1,4)& (35,75)  & (900,2925)  & (90005,428625)              \\
(1,5)& (56,126) & (2451,8436) &                             \\
\hline
(2,1)& (20,40)     & (90,240)        & (544,1992)           \\
(2,2)& (56,126)    & (705,2280)      & (19548,88452)        \\
(2,3)& (120,288)   & (3400,11968)    & (306016,1468480)     \\
(2,4)& (220,550)   & (12025,44200)   & (2725010,13350750)   \\
(2,5)& (364,936)   & (34410,129648)  &                      \\
\hline
(3,1)& (56,126)    & (357,1092)      & (3258,13338)         \\
(3,2)& (165,405)   & (3132,10962)    & (137268,645408)      \\
(3,3)& (364,936)   & (15900,58800)   & (2249184,10964880)   \\
(3,4)& (680,1800)  & (57750,219450)  &                      \\
(3,5)& (1140,3078) & (167805,647460) &                                       
    \end{tabular}
    \caption{Numbers of non-isomorphic graphs/orbits of type at most $(n,v_c,e_c)$.}
    \label{tab1}
\end{table}

Isomorphism naturally results in canonical labels, an invariant among isomorphic graphs. Our orbit generation problem can be generalized to a graph plus an ordered partition. Let $\Pi$ denote the set of ordered partitions on $V$. Then the action $g$ on $V$ induces a new action on $\Pi$ by $\pi^g=(W_1,\dots,W_m)^g=(W^g_1,\dots,W^g_m)=(\{v^g:v\in W\})_{W\in\pi}$ for $\pi=(W_1,\dots,W_m) \in\Pi$. Two pairs $(M,\pi),(M',\pi')$ are \emph{isomorphic} if there is a $g\in S_n$ such that $(M,\pi)^g=(M^g,\pi^g)$. \\
A \emph{canonical label} is a function
\begin{equation}
    C:\mathcal G\times \Pi\to \mathcal G\times \Pi
\end{equation}
such that, for all $M\in \mathcal G, \pi\in \Pi$ and $g\in S_n$,
\begin{enumerate}
    \item[(C1)] $C(M,\pi)\simeq (M,\pi)$ 
    \item[(C2)] $C(M^g,\pi^g)=C(M,\pi)$.
\end{enumerate}
Note that $(M,\pi)\simeq (M',\pi')\iff C(M,\pi)\simeq C(M',\pi')$. \\

The \textit{equitable partitioning} technique is crucial for McKay's algorithm \cite{mckay1981practical}. It can be extended 
 to canonical labels of colored graphs. Let $v\in V$ and $\pi=(W_1,\dots,W_m)\in\Pi$. The \emph{degree of $v$ in $W_i$}, denoted by $\deg_{W_i}(v)$, is the sum of the entries at the intersection of the row of adjacency matrix $A_M$ corresponding to $v$ and columns corresponding to $W_i$. The ordered partition $\pi$ is \emph{equitable} if 
\begin{equation}
    \deg_{W_i}(v)=\deg_{W_i}(w)\,\,\, \forall v,w \in W_j
\end{equation}
holds for all $i,j\in\{1,\dots,m\}$. \\

Based on equitable partition, \cite{mckay1981practical} provided an algorithm  to compute the permutation required to arrange $V$ to obtain the canonical labels for simple graphs. That is, $\pi^M_{lab}$ is a discrete order partition, i.e., a permutation such that
$(M,\pi)^{\pi^M_{lab}}$ is the canonical label of $(M,\pi)$. Although their proof of the algorithm's correctness is for simple graphs, it also works for heterogeneous graphs. \\

\begin{figure}[ht]
    \centering
    \includegraphics[scale=0.5]{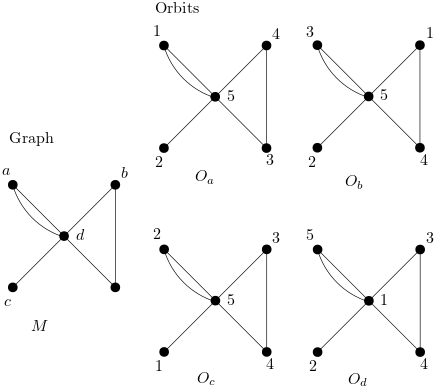}
    \caption{Graph and orbits of a heterogeneous graph with type $(n, v_c, e_c) = (5,1,2)$. }
    \label{fig1}
\end{figure}

Using this tool, the vertex set $\mathcal{O}_{v,M}$ can be represented by a unique graph, with vertex $1$ separated from all other vertices in the ordered partition $\pi$ of $V$
$$
O_v^{\pi^{M_v}_{lab}},
$$
where $O_v$ is the graph obtained from $M$ by swapping vertices $v$ and $1$, and $\pi=(\{1\}, V-1)$. More formally,
$$
\mathcal{O}_{v,M}=\{u:O_u^{\pi^{O_u}_{lab}}=O_v^{\pi_{lab}^{O_v}}\}.
$$
In other words, $O_v$ is the canonical form when the first vertex is reference point $v$.
Figure \ref{fig1} shows an example for type $(n, v_c, e_c) = (5,1,2)$, where there are four orbits relative to this graph. 
Therefore, we obtain four graphs $O_a, O_b, O_c, O_d$ corresponding to the four orbitals $O_{a,M}, O_{b,M}, O_{c,M}, O_{d,M}$, respectively. Vertex $1$ in $O_a$ is vertex $a$ in $M$. The remaining vertices, say $e$, are in orbit $b$. Therefore, the graph it produces is $O_b$; that is, $O_e=O_b$.

Using other vertices to assign the graphs gives the same result. Thus, the first vertex is selected for convenience.

\section{Proposed package} 
This package was written in Python 3 with the NumPy library. 
The three entry functions are \code{(Di)Canonical}, \code{(Di)Generator}, and \code{(Di)Count}, which can be used to distinguish graphlet orbits, generate a list of non-isomorphic heterogeneous graphs and orbits, and return a vector consisting of the numbers of occurrences of graphs and orbits of a given graph, respectively. \\

\code{(Di)Canonical(G, ref = -1)}:
This function computes the invariant of isomorphic graphs or graphlet orbits, which can be used to distinguish non-isomorphic graphs or graphlet orbits. \\

If \code{ref} $<1$, the function computes an \textit{invariant} of graphs isomorphic to the input graph \code{G}, e.g., $G_1$ in Figure~\ref{fig2}. Otherwise, it computes an invariant of the graphlet orbit associated with $\mathcal{O}_{ref,G}$ defined in Section~\ref{sec: algorithm}, e.g., graphlet orbit $3$ in Figure~\ref{fig2}.\\

Input:
An adjacency matrix \code{G} representing a graph and, optionally, a vertex \code{ref} $\geq 0$.
\subsection{Output}
The serialized adjacency matrix of the canonical label $C(G)$ of the input graph \code{G}, or if vertex \code{ref} $\geq 0$, the serialized adjacency matrix of
$$
M_{ref}^{\pi_{lab}^{M_{ref}}}
$$
defined in Section~\ref{sec: algorithm}.
Implementation:
If \code{ref} $<1$, then it returns the lexicographically smallest serialized adjacency matrix of \code{G} for all $|V(G)|!$ permutations of its vertices. Otherwise, the algorithm swaps vertex \code{0} with vertex \code{ref} and returns the lexicographically smallest serialized adjacency matrix for all $|V(G)-1|!$ permutations of vertices $1,2,3,\ldots,|V(G)|-1$. When comparing the serialized adjacency matrix for undirected graphs, we only need to consider the entries on or above the diagonal because the adjacency matrix is symmetric. With directed graphs, the entire adjacency matrix is required. \\

\code{(Di)Generator(n, sizev, sizee, orbg = False, connect = False, Time = True)}:

Built upon \code{(Di)Canonical}, this function generates a list of non-isomorphic graphs of order \code{n} or graphlet orbits from graphs of order \code{n}, where each vertex has a color in [\code{sizev}] and each edge has a color in [\code{sizee}].

Input:
\begin{itemize}
\item[\code{n}]:
The order of graphs to be considered.
\item[\code{sizev}]:
Maximum number of vertex colors.
\item[\code{sizee}]:
Maximum number of edge colors.
\item[\code{orbg}]:
Whether to generate graphs (\code{False}) or graphlet orbits (\code{True}). Default is \code{False}.
\item[\code{connect}]:
Whether to restrict to connected graphs (\code{True}) or not (\code{False}). Default is \code{False}.
\item[\code{Time}]:
Whether to log the running time (\code{True}) or not (\code{False}). Default is \code{True}.
\end{itemize}
Output: The list of adjacency matrices representing the non-isomorphic graphs or graphlet orbits.

Implementation: The function uses Dynamic Programming. It recursively calls itself with a smaller number of vertices, vertex colors, and edge colors to gradually obtain the intended list from the base cases, where graph order and numbers of edge and vertex colors are small. See Section~\ref{section:graph enumeration} and Section~\ref{section:orbit enumeration from a graph} for more details. It uses \code{(Di)Canonical} to determine whether two graphs or graphlet orbits are isomorphic.
\\
\\
\code{(Di)Count(G, ref, k, sizev, sizee, connect = False}): Given a host graph \code{G} and its vertex \code{ref}, this function returns a vector consisting of the number of occurrences, at \code{ref} in \code{G}, of graphlet orbits belonging to graphs of order \code{k}, where each vertex has a color in [\code{sizev}] and each edge has a color in [\code{sizee]}.

Input:
\begin{itemize}
\item[\code{G}]:
The host graph in which the occurrences of graphlet orbits are to be counted.
\item[\code{ref}]:
A vertex in \code{G} at which the function counts the number of occurrences of the specified graphlet orbits.
\item[\code{k}]:
The function counts in \code{G} the occurrences of all graphlet orbits belonging to graphs of order \code{k}.
\item[\code{sizev}]:
Maximum number of vertex colors in graphs of order \code{k} in which the graphlet orbits are considered.
\item[\code{sizee}]:
Maximum number of edge colors in graphs of order \code{k} in which the graphlet orbits are considered.
\item[\code{connect}]:
Whether to restrict to connected graphs (\code{True}) or not (\code{False}). Default is \code{False}.
\end{itemize}
Output: A vector containing the number of occurrences of all graphlet orbits considered.

Implementation: The function first finds all graphlet orbits of interest via \code{(Di)Generator}. The number of occurrences of these graphlet orbits is then calculated. \\

\subsection{Examples}
To illustrate our package, we provide four examples containing both synthetic and real-world data. 

\begin{example}
Let 
\begin{equation}
    M=\left(\begin{array}{cccccc}
0& 1& 1& 0& 0& 0\\
1& 0& 1& 0& 0& 0\\
1& 1& 0& 1& 1& 0\\
0& 0& 1& 0& 0& 0\\
0& 0& 1& 0& 0& 1\\
0& 0& 0& 0& 1& 0
\end{array}\right).
\end{equation}
Then,  \code{Canonical(M)} will return the list \code{[0, 0, 0, 0, 1, 0, 0, 0, 0, 0, 0, 0, 0, 1, 0, 1, 0, 1, 1, 1, 0]}.
The matrix can be recovered from the list using \code{lis2mat(L,6)}. Finally, we obtain a graph under canonical labeling. We also treat $M$ as a digraph. Then,  \code{DiCanonical(M)} will return the list \code{[0, 0, 0, 0, 0, 0, 0, 0, 0, 0, 0, 1, 0, 1, 0, 0, 0, 1, 0, 0, 0, 0, 0, 1, 0, 1, 0, 1, 1, 1, 0, 1, 1, 1, 0, 1]}. 

Again, \code{lis2mat(L,6)} returns a graph under canonical labeling. Note that we get different canonical labeling for graphs and digraphs; refer to Figure \ref{fig:cano}. To compute the canonical form of orbit $O_{5,M}$, we enter \code{Canonical(M,4)}. 

\begin{figure}[ht]
    \centering
    \includegraphics[scale=0.5]{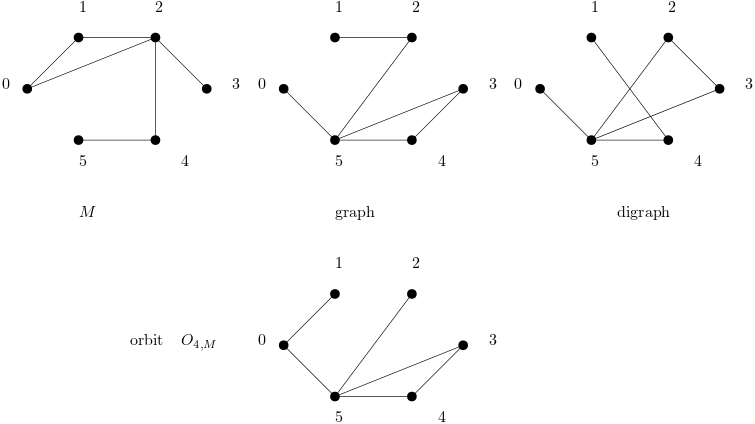}
    \caption{Canonical labeling for graphs and digraphs}
    \label{fig:cano}
\end{figure}

\end{example}

\begin{example}
If we enter \code{Generator(4,1,1)}, it returns a list of all graphs of type at most $(n,v_c,e_c) = (4,1,1)$. Thus, we get
\code{[[0, 0, 0, 0, 0, 0, 0, 0, 0, 0],
       [0, 0, 0, 0, 0, 0, 0, 0, 1, 0],
       [0, 0, 0, 0, 0, 0, 0, 1, 1, 0],
       [0, 0, 0, 0, 0, 0, 1, 1, 1, 0]\\,\dots]}.
Two parameters can be changed: \code{orbg} and \code{connect}.
Thus, the command
\code{Generator(4,\\1,1,orbg=True,connect=True)} will return the list of all connected orbits of type at most $(n,v_c,e_c) = (4,1,1)$. Therefore, we obtain
 \code{[0, 0, 0, 0, 0, 0, 1, 1, 1, 0],
       [0, 0, 0, 0, 1, 0, 1, 1, 1, 0],
       [0, 0, 0, 1, 0, 0, 0, 1, 1, 0],
       [0, 0, 0, 1, 0, 0, 1, 1, 1, 0],\dots]}.
\end{example}

\begin{example}[Dolphin network]

The dolphin network was used to assess the frequency at which two dolphins were observed together. The sexes of all but four dolphins were known. Whenever a dolphin pod was spotted in a particular fjord between 1995 and 2001, each adult member of the pod was photographed and identified based on the natural markings on its dorsal fin \cite{lusseau2003emergent}. The whole graph is presented in \ref{dolph}. The network is a heterogeneous graph of type $(n,v_c,e_c) = (62,3,1)$. The orbits that second dolphin appears in are listed in Table \ref{tab2}.

\begin{figure}[ht]
    \centering
    \includegraphics[scale=0.4]{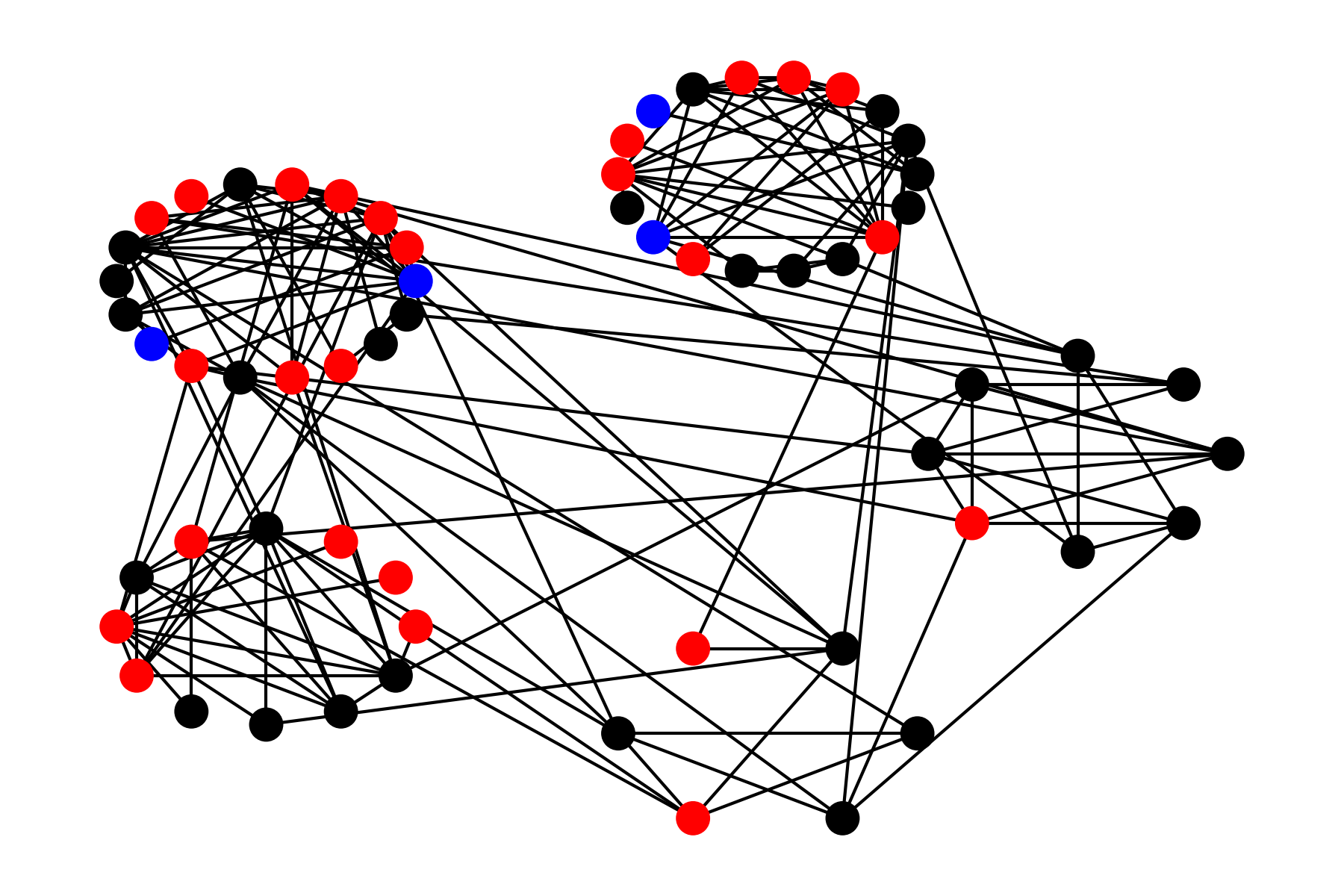}
    \caption{Dolphin network: The structure of the whole dolphin network. Red (blue, black) vertices indicate that the dolphins are female (male, unknown).}
    \label{dolph}
\end{figure}
Let $M$ be the adjacency matrix of the dolphin network.
If we enter \code{Count(M, 1, 4, 3, 1, True)}, it will return the statistical data for the second dolphin (Python starts from 0). This is a list of length 627.
Figure \ref{dol} shows Orbits 0, 4, 6, 13, and 24.
\begin{figure}[ht]
    \centering
    \includegraphics[scale=0.4]{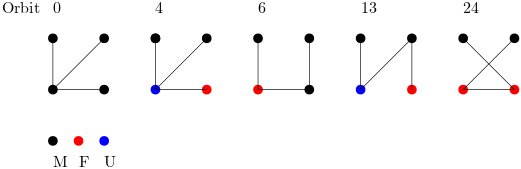}
    \caption{Some orbits of the dolphin network. The second dolphin is in the upper left.}
    \label{dol}
\end{figure}

\begin{table}
    \centering
    \begin{tabular}{c|c}
     frequency    & Orbit  \\
   1 & 2, 7, 8, 18, 21, 23, 27, 29, 35, 41, 43, 49, 50, 52, 59, 66,\\
   & 74, 75, 95, 99, 105, 118, 140, 142, 153, 154, 181, 186\\
   2&0, 4, 6, 13, 24\\
   3&3, 124\\
   4&90, 93, 126, 136, 138\\
   5&63, 107, 123, 135\\
   6&38\\
   7&121\\
   9&33\\
   10&131\\
   12&19\\
   14&37\\
   15&93\\
   18&122\\
   20&116
    \end{tabular}
    \caption{Numbers of occurrences}
    \label{tab2}
\end{table}

\end{example}

\begin{example}
For illustrative purposes, we constructed an artificial heterogeneous graph. The directed network in Figure \ref{resource} has 10 vertices, 3 types of nodes, and 2 types of directed edges. We now count the frequency of vertex 0 as an orbit of type $(n,v_c,e_c) = (4,3,2)$. Therefore, we enter \code{DiCount(M, 0, 4, 3, 2, True)}. There are 7161912 connected orbits, and only 84 orbits occur all with frequency 1. Figure \ref{diorb} shows the first five orbits: 5145192, 5145203, 5145456, 5145776, and 5146300.

\begin{figure}[ht]
    \centering
    \includegraphics[scale = 0.3]{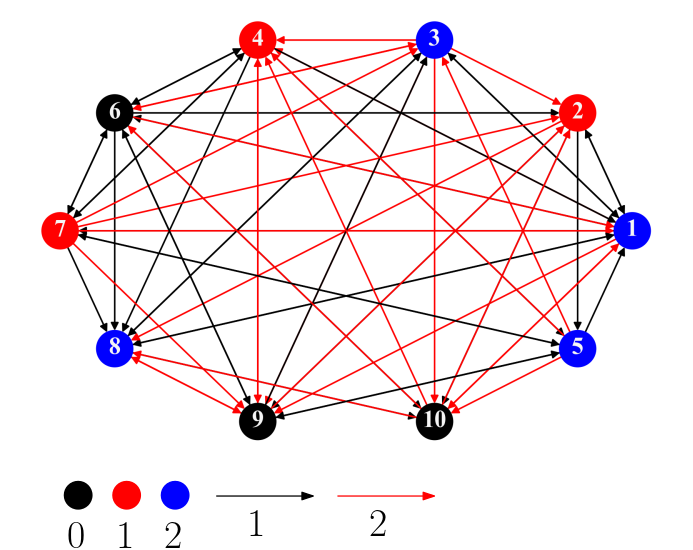}
    \caption{A heterogeneous network with directed edges.}
    \label{resource}
\end{figure}

\begin{figure}[ht]
    \centering
    \includegraphics[scale=0.4]{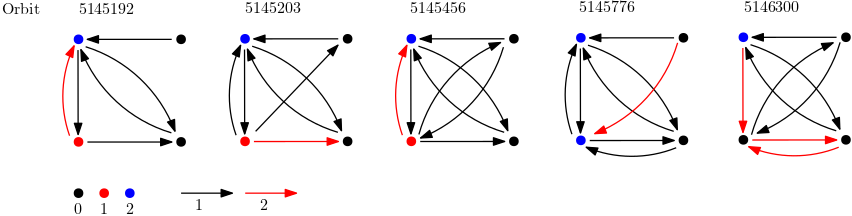}
    \caption{First five orbits of directed graph.}
    \label{diorb}
\end{figure}

\end{example}

\begin{example}[New example]
\begin{equation}
M=  \left(  \begin{array}{ccccccc}
0&1&0&0&0&0&0\\
0&0&1&0&1&1&0\\
0&0&0&1&0&0&0\\
1&0&0&0&0&0&1\\
0&0&0&0&0&0&0\\
0&0&0&0&0&0&0\\
0&0&1&0&0&0&0\end{array}\right)\end{equation}

 The directed network in Figure \ref{resource} has 7 vertices, 1 types of nodes, and 1 types of directed edges.
 The \code{DiCanonical(M)} gives \code{[0, 0, 0, 0, 0, 0, 0, 0, 0, 0, 0, 0, 0, 0, 0, 0, 0, 0, 0, 0, 0, 0,
       0, 0, 1, 0, 0, 0, 1, 0, 0, 0, 0, 1, 1, 1, 1, 0, 0, 0, 0, 0, 0, 0,
       1, 1, 0, 0, 0]}
 
 We now count the frequency of vertex $2$ as an orbit of type $(n,v_c,e_c) = (4,1,1)$. Therefore, we enter \code{DiCount(M, 0, 4, 1, 1, True)}.


\begin{figure}[ht]
    \centering
    \includegraphics[scale = 0.3]{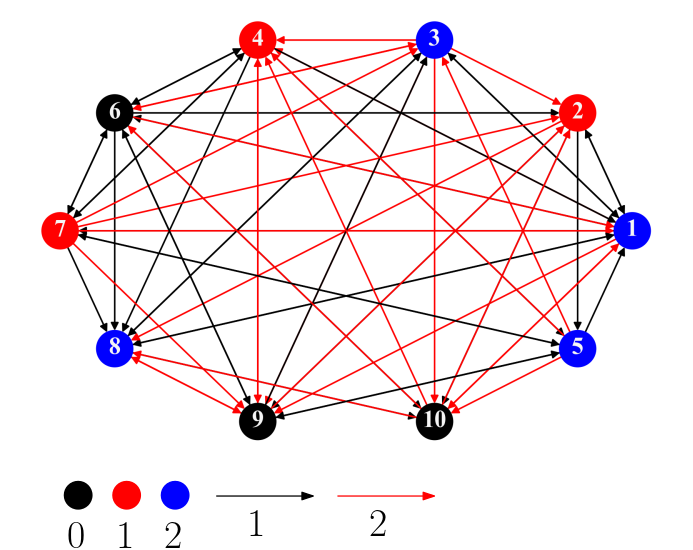}
    \caption{A heterogeneous network with directed edges.}
    \label{resource}
\end{figure}

\begin{figure}[ht]
    \centering
    \includegraphics[scale=0.4]{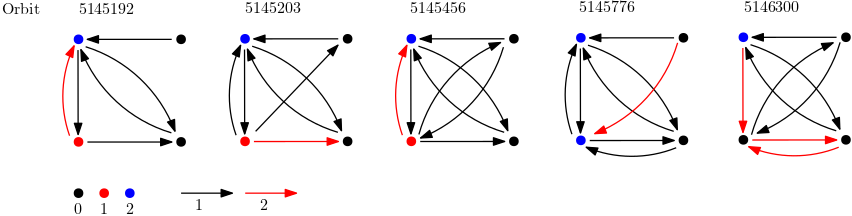}
    \caption{First five orbits of directed graph.}
    \label{diorb}
\end{figure}

\end{example}

\section{Conclusion}
Graphlets provide an efficient topological representation of how each node and edge is integrated inside the mesoscale structure of a network and are widely utilized in data science. In this study, we built a Python package to generate a canonical form of graphlets and to count the occurrences of graphlet orbits in heterogeneous graphs. By generalizing the techniques for distinguishing graphlet orbits and generating a list of non-isomorphic graphs and orbits for heterogeneous graphs, we made possible sophisticated descriptions of directed and directed complex networked data that were not possible with single-type graphlets. Once the order and type are calculated, the information can be saved, and the program will directly use the file the next time. In the package, we count the frequency of some special subgraphs, but not all subgraphs, by counting orbits to solve a linear system of equations. Therefore, our package can save considerable time for researchers analyzing graphlets and orbits of heterogeneous graphs. However, the number of non-isomorphic graphs/orbits increases exponentially with type and order. The size of the file that stores all the canonical forms also increases significantly as the number of parameters increases. For example, the file size for directed graph type $(n,v_c,e_c) = (4,3,2)$ is approximately 0.8 GB with 1426032 orbits. Because of the essentially exponential-ordered complexity of the enumerations of graphlets and orbits, our package is more suitable for heterogeneous graphs with small order, less node type, and less edge type parameters, which can be reused many times. A novel contribution of this study is the use of a generating method to count the orbits of edges. Consequently, the analytical methods and computing techniques used in this study can potentially yield new discoveries in various heterogeneous graphs in various fields. Our package is available at \cite{pypi} and users are encouraged to contribute to it. 

\section{Acknowledgements}
Shen-Fu Tsai is supported by the Ministry of Science and Technology of Taiwan under grant MOST 111-2115-M-008-010-MY2.

\bibliography{ref}

\end{document}